\pdfoutput=1
\documentclass[11pt]{article}

\usepackage{acl}
\usepackage{enumitem}
\usepackage{times}
\usepackage{latexsym}

\usepackage[T1]{fontenc}

\usepackage[utf8]{inputenc}
\usepackage{times}  
\usepackage{helvet}  
\usepackage{courier}  
\usepackage{booktabs}
\usepackage{rotating}
\usepackage{graphicx} 
\usepackage{natbib}  
\usepackage{multicol}
\usepackage{multirow}
\usepackage{adjustbox}
\usepackage{algpseudocode}
\usepackage{caption} 

\usepackage{algorithm}
\usepackage{caption}
\usepackage[]{color-edits}
\usepackage{newfloat}
\usepackage{listings}
\DeclareCaptionStyle{ruled}{labelfont=normalfont,labelsep=colon,strut=off} 
\floatstyle{ruled}
\newfloat{listing}{tb}{lst}{}
\floatname{listing}{Listing}

\usepackage{microtype}

\title{RecMind: Large Language Model Powered Agent For Recommendation}

\author{
    Yancheng Wang$^{1}$\thanks{~~Work was done as an intern at Amazon Alexa AI.},
    Ziyan Jiang$^{2}$\thanks{~~Indicates equal contribution.},
    Zheng Chen$^{2}$\footnotemark[2],
    Fan Yang$^{2}$\footnotemark[2],
    Yingxue Zhou$^{2}$\footnotemark[2], \\
    {\bf Eunah Cho}$^{2}$,
    {\bf Xing Fan}$^{2}$,
    {\bf Xiaojiang Huang}$^{2}$,
    {\bf Yanbin Lu}$^{2}$,
    {\bf Yingzhen Yang}$^{1}$ \\
        $^1$School of Computing and Augmented Intelligence, Arizona State University\\
    $^2$Amazon Alexa AI\\
    { \{yancheng.wang, yingzhen.yang\}@asu.edu}\\
    { \{ziyjiang, zgchen, ffanyang, zyingxue, eunahch, fanxing, xjhuang, luyanbin\}@amazon.com}
}
\begin{document}
\maketitle

\begin{abstract}
While the recommendation system (RS) has advanced significantly through deep learning, current RS approaches usually train and fine-tune models on task-specific datasets, limiting their generalizability to new recommendation tasks and their ability to leverage external knowledge due to model scale and data size constraints. Thus, we designed an LLM-powered autonomous recommender agent, RecMind, which is capable of leveraging external knowledge, utilizing tools with careful planning to provide zero-shot personalized recommendations. We propose a \textit{Self-Inspiring} algorithm to improve the planning ability. At each intermediate step, the LLM “self-inspires” to consider all previously explored states to plan for the next step. This mechanism greatly improves the model’s ability to comprehend and utilize historical information in planning for recommendation. We evaluate RecMind's performance in various recommendation scenarios.
Our experiment shows that RecMind outperforms existing zero/few-shot LLM-based recommendation baseline methods in various tasks and achieves comparable performance to a fully trained recommendation model P5. 
\end{abstract}

\section{Introduction}
The \textit{Recommender System} (RS) plays a key role in search engines, e-commerce,
and various other Internet platforms. An RS analyzes the historical interactions between users and items to recommend potential items
\citep{Koren2009MatrixFT,Linden2003AmazoncomRI}. 
The RS has been enhanced by Deep Neural Networks (DNNs) to more effectively learn the representations of users, items, and sequential behaviors \citep{hidasi2015session, he2020lightgcn, sun2019bert4rec}. However, most existing DNN-based methods (e.g., CNN and LSTM) and pre-trained language models (e.g., BERT) cannot sufficiently capture textual knowledge about users and items due to limitations in model scale and data size. Besides, most existing RS methods have been designed for specific tasks and are inadequate in generalizing to unseen recommendation tasks \citep{fan2023recommender}. 

Recent advances in Large Language Models (LLMs), such as GPT-3 \citep{brown2020language}, GPT-4 \citep{openai2023gpt}, LLaMA \citep{touvron2023llama}, LLaMa-2 \citep{touvron2023llama2}, and PaLM-2 \citep{anil2023palm} have demonstrated remarkable results in a wide range of tasks, which have motivated the research of leveraging LLMs for recommendation to mitigate the aforementioned challenges \citep{Liu2023IsCA, fan2023recommender, Lin2023HowCR}. However, existing studies primarily rely on knowledge stored within the model's weights, neglecting the potential benefits of leveraging external tools to access real-time information and external knowledge \citep{yang2023palr,bao2023tallrec}. Furthermore, the reasoning ability of LLMs is not fully utilized for recommendation, resulting in suboptimal predictions due to the intricate nature of recommendation-related tasks \citep{Liu2023IsCA}.


\begin{figure*}[ht]
    \centering
    \includegraphics[width=\textwidth]{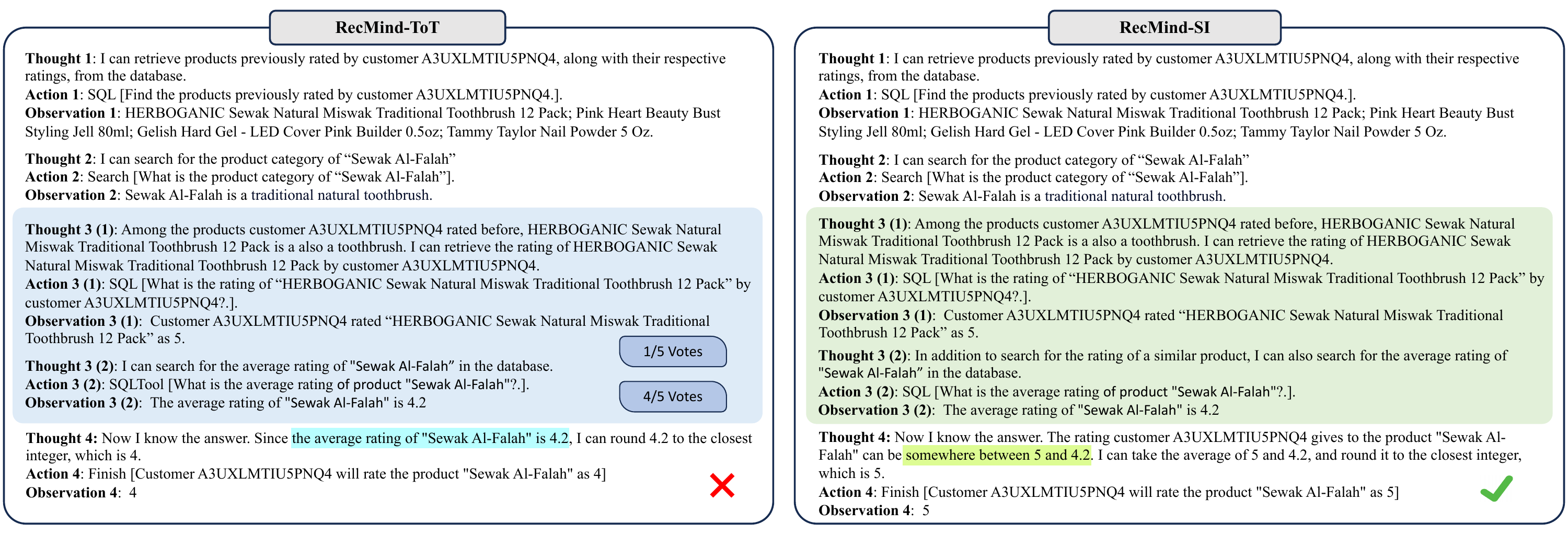}
    \caption{Comparisons of rating prediction by RecMind-ToT (left) and RecMind-SI (right). After searching for the product category of the item in Step 2, RecMind-ToT first generates thought 3 (1) to retrieve the rating of a similar item. After being evaluated by the voting-based evaluator,  RecMind-ToT prunes option 3 (1) and proposes another thought 3 (2) to retrieve the average rating of the item and then makes the prediction solely based on it. In contrast, although RecMind-SI proposed the same alternative options in step 3, it takes into account the thought, action, and observation from both options 3 (1) and 3 (2) to generate the thought for the next step.} 
    \vspace{-5mm}
    \label{fig:example}
\end{figure*}



To better utilize the strong reasoning and tool-using abilities of LLMs, we design a recommendation agent RecMind that leverages an LLM-powered API as its intellectual core and incorporates a few key components. The first key component is \textbf{Planning} which enables the agent to break complex recommendation tasks into manageable steps for efficient handling of complex situations. Each step of planning involves \textit{thought}, \textit{action} and \textit{observation} (see Figure \ref{fig:example} for examples and Section \ref{sec:architecture} for details). The agent is also equipped with \textbf{Memory} consisting of \textit{Personalized Memory} and \textit{World Knowledge}, each accessible through specific tools. The \textbf{Tools} enhance the agent's functionality on top of the LLM, such as retrieving relevant knowledge, or assisting with the reasoning process. 

To further enhance the planning ability of the agent, we propose a new planning algorithm \textit{Self-Inspiring} (SI). At each intermediate planning step, the agent “self-inspires” to consider all previously explored paths for the next planning. Unlike existing Chain-of-Thoughts (CoT) \citep{Wei2022ChainOT} and Tree-of-Thoughts (ToT) \citep{Yao2023TreeOT} which discards states (thoughts) in previously explored paths when generating a new state, SI retains all previous states from all history paths when generating new state. SI is inspired by the intuition that all historical states can provide useful information for better planning. Figure \ref{fig:example} provides an example of ToT and SI showing that SI achieves a more accurate rating than ToT due to better planning.

To the best of our knowledge, this is the first public research work on an LLM-powered autonomous agent for recommendation. The main contributions of our work are:\vspace{-1mm}
\begin{itemize}[leftmargin=5mm]
    \item We introduce RecMind, the first LLM-powered agent designed for general recommendation purposes, which operates without the need for fine-tuning for domain adaptation across datasets or tasks. \vspace{-1mm}
    \item We incorporate a novel \textit{self-inspiring} (SI) planning technique in RecMind. 
    This technique integrates multiple reasoning paths and offers an empirical improvement over currently popular methods, such as CoT and ToT. \vspace{-1mm}
    \item We evaluate the effectiveness and generalizability of RecMind across five recommendation tasks and two datasets. Extensive experiments and analyses demonstrate that RecMind outperforms state-of-the-art (SOTA) LLM-based baselines that do not involve any fine-tuning and achieves competitive performance with a fully pre-trained expert recommendation model such as P5 \citep{geng2022recommendation}. In addition, SI outperforms CoT and ToT on general reasoning tasks, showing that the proposed the impact of SI is beyond recommendation tasks.
\end{itemize}

\begin{figure*}[!t]
    \centering
    \vspace{-2mm}
    \includegraphics[width=0.95\textwidth]{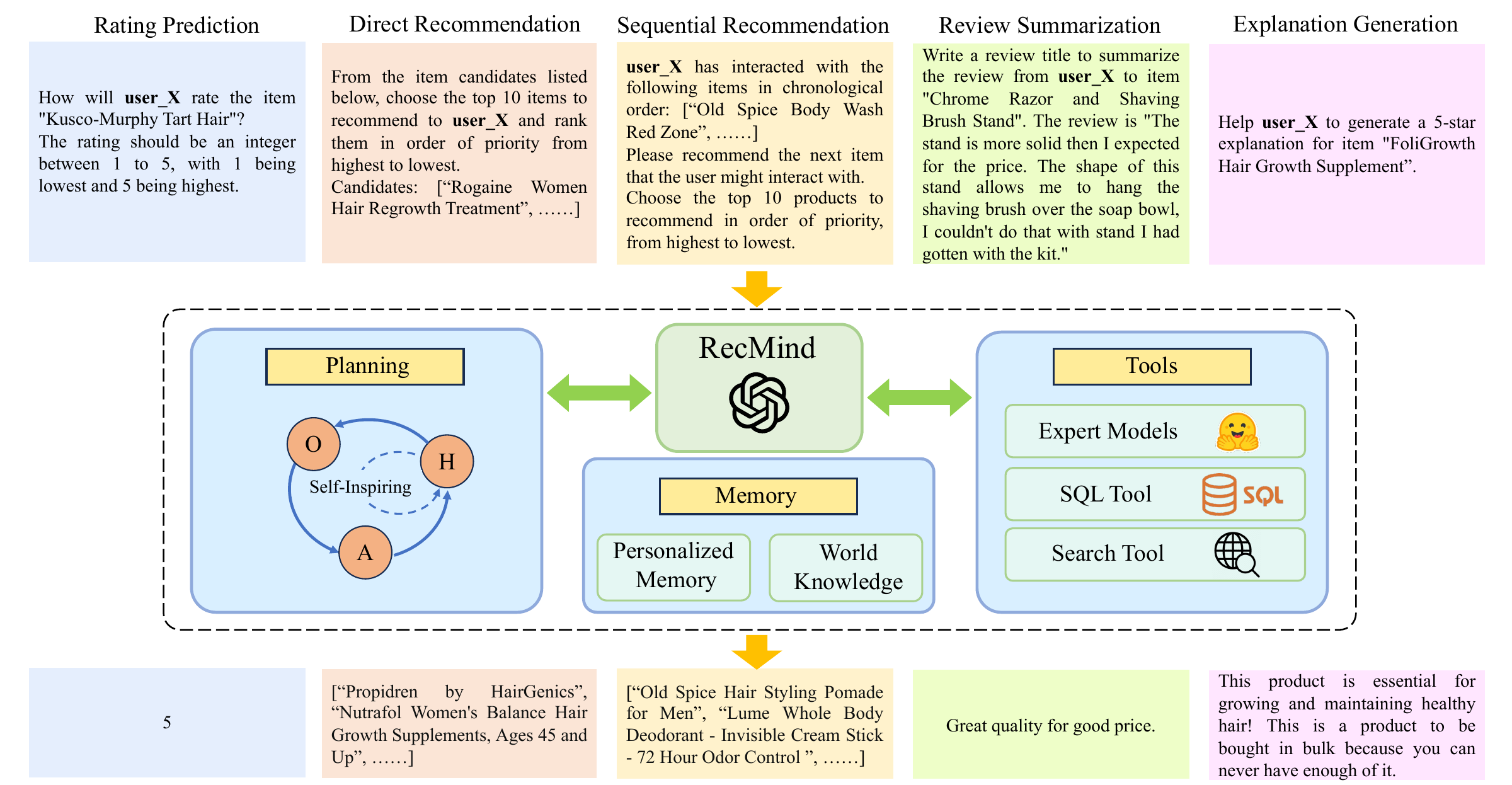}
    \vspace{-2mm}
    \caption{Here is an overview of our proposed RecMind architecture. It comprises four major components: "RecMind" is built based on ChatGPT API, "Tools" support various API call to retrieve knowledge from "Memory" component, "Planning" component is in charge of thoughts generation.} 
    \vspace{-2mm}
    \label{fig:architecture}
\end{figure*}

\vspace{-3mm}
\section{Related Work}
\noindent\textbf{LLM-as-Agent} There is an emerging trend where LLMs are augmented to become autonomous language agents. 
The central concept is to leverage LLMs to produce text-based outputs and actions that can be used for making API calls and performing operations within a specific environment. LLMs, with their strong reasoning abilities, can decompose challenging and complex tasks into smaller, more manageable steps \citep{Wei2022ChainOT, Yao2023TreeOT, patil2023gorilla}. 
A number of successful applications have emerged, including ReAct \citep{yao2022react}, Toolformer \citep{Schick2023ToolformerLM}, HuggingGPT \citep{shen2023hugginggpt}, generative agents \citep{park2023generative}, WebGPT \citep{nakano2021webgpt}, AutoGPT \citep{SignificantGravitas2023}, BabyAGI \citep{nakajima2023}, and Langchain \citep{langchainai2023}.

\noindent\textbf{LLM for Recommendation} Recently, LLMs have gained popularity in recommender systems, given their ability to understand a user's preferences or past interactions in natural language \citep{fan2023recommender, Lin2023HowCR}. Current LLM-based recommender systems are primarily designed for rating prediction \citep{kang2023llms, bao2023tallrec} and sequential recommendation tasks \citep{Wang2023ZeroShotNR, yang2023palr, hou2023large}. 
User interactions and optional data, including profiles, are input into an LLM prompt with options: no fine-tuning \citep{Wang2023ZeroShotNR}, full-model fine-tuning \citep{yang2023palr}, or parameter-efficient fine-tuning \citep{bao2023tallrec}. In sequential recommendation tasks, we use a pre-filtered set of item candidates in the prompts for focused ranking.
\cite{Liu2023IsCA} use prompts to assess ChatGPT's performance across five recommendation tasks showing LLM's strong in-context learning abilities and generalization \citep{wei2021finetuned}. Unlike existing studies, our work 
harnesses the LLM's capabilities in reasoning, tool usage, and action. 

\section{Architecture}
\label{sec:architecture}
As shown in Figure \ref{fig:architecture}, RecMind consists of key components: LLM-powered API such as ChatGPT to drive the overall reasoning, \textbf{planning} which breaks down a task to smaller sub-tasks for step-by-step planning, \textbf{memory} which provides the agent with the capability to retain and recall information over extended periods, and \textbf{tools} for obtaining relevant extra information from memory that is missing from the model weights and aiding the reasoning. We introduce the key components \textbf{planning}, \textbf{memory} and \textbf{tools} for RecMind in the subsequent parts.

\begin{figure}[t]
    \centering
    \includegraphics[width=\columnwidth]{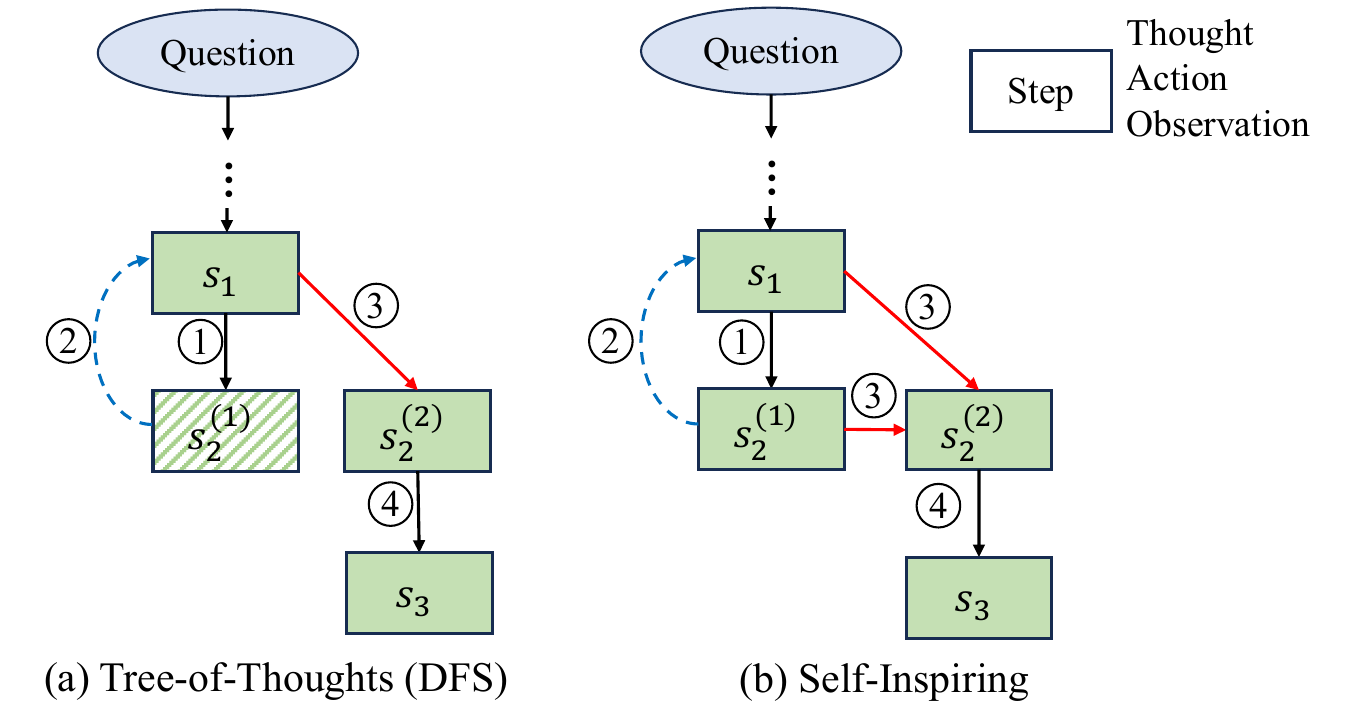}
    \vspace{-7.5mm}
    \caption{Comparison between Tree-of-Thoughts DFS and Self-Inspiring. Red arrows in the figure indicate the process for generating alternative thoughts at intermediate steps.  Blue dashed arrows in the figure denote the backtracking process.} 
    \vspace{-5mm}
    \label{fig:self_inspiring}
\end{figure}

\paragraph{Planning} Planning helps LLM Agents decompose tasks into smaller, manageable subgoals for handling complex tasks. Consider the setting where the goal is to generate the final result $y$ given problem $x$ via an LLM Agent parameterized by $\theta$. The traditional input-output method gives the result by $y \sim p_\theta(y|x)$. With planning, RecMind generates the result $y \sim p_\theta(y|\texttt{planing}(x))$, where $\texttt{planing}(x)$ is a set of prompts that decomposes problem $x$ into a series sub-tasks that is composed of t\textbf{h}ought $h$, \textbf{a}ction $a$, and \textbf{o}bservation $o$. Figure \ref{fig:example} provides examples of planning including thoughts, actions, and observations. We first review existing popular reasoning methods such as CoT and ToT which we have explored for RecMind. Then we present the proposed  SI algorithm. All these planning methods can be viewed as traversing through a latent \textit{reasoning tree}, as shown in Figure \ref{fig:self_inspiring}.
\vspace{-1mm}
\begin{itemize}[leftmargin=*]
    \item \textbf{Chain-of-Thoughts (CoT)} \citep{Wei2022ChainOT} has been used in ReAct \citep{yao2022react} to synergize reasoning and action. This CoT planning method follows a single path in the reasoning tree. In our setting, at each time step $t$, the agent receives observation $o_t$ followed by thought $h_t$ and action $a_t$. Let $s_t =(h_t, a_t, o_t)$ denote the RecMind state at step $t$. The CoT planning method generates the next state $s_{t+1} =(h_{t+1}, a_{t+1}, o_{t+1})$ by sampling $p_\theta(s_{t+1}|x, s_1, .., s_t)$. Thus CoT only follows a single planning path $S = \{s_1, ..., s_t, ..., s_T\}$ until reaching the final result $y \sim p_\theta(y | x, s_1, ..., s_t,..., s_T)$ after $T$ steps.
    \vspace{-1mm}
    \item \textbf{Tree-of-Thoughts (ToT)} \citep{Yao2023TreeOT} extends CoT to explore multiple paths in the reasoning tree. At step $t$ and state $s_t$, ToT-BFS explicitly generates multiple candidates $\{s^1_{t+1}, ..., s^k_{t+1}\}$ for next state by i.i.d. sampling $s^i_{t+1} \sim p_\theta(s_{t+1}|x, s_1, .., s_t)$ for $i \in [k]$. Then it applies majority vote to select the state $s_{t+1}$ from $\{s^1_{t+1}, ..., s^k_{t+1}\}$. Eventually ToT-BFS generates a single path similar to CoT. In contrast, ToT-DFS explores one branch at a time, but might prune the current state, and backtracks to the previous state to start a new reasoning branch. Denote the first explored path as $z^{(1)} = \{s_1^{(1)}, ..., s_t^{(1)}, s^{(1)}_{t+1}\}$. If the last state $s^{(1)}_{t+1}$ is pruned and it backtracks to the previous state $s^{(1)}_{t}$, and starts a new reasoning branch, then the path becomes $z^{(2)} = \{s_1^{(1)}, ..., s_t^{(1)}, s^{(2)}_{t+1}, ...\}$. After exploring $n$ branches, we denote the final path of ToT as $z^{(n)} = \{s_1, ..., s^{(1)}_{j_1}, ..., s^{(2)}_{j_2}, ..., s^{(n)}_{T}\}$ and the final result $y$ is obtained by $y \sim p_\theta(x, z^{(n)})$. 
\end{itemize}

We find the discarded historical states from previously explored branches such as $s^{(1)}_{t+1}$ from branch $z^{(1)}$ usually contain helpful information for RecMind to generate a better state compared with only considering the final path of ToT. Thus, we propose \textbf{\textit{Self-Inspiring} (SI)} as shown in Figure \ref{fig:self_inspiring}(b) and Algorithm 1, a new planning method for RecMind. SI \textit{inspires} itself into exploring an alternative reasoning branch, while retaining all previous states. At $m$-th path and step $t$, SI generates the next step of planning by considering all previous paths, i.e., $s^{(m)}_{t+1} \sim p_\theta(s_{t+1}|z^{(1)}, ..., z^{(m)})$. After exploring $n$ paths, the RecMind obtains the final result $y \sim P_\theta(x, z^{(1)}, ..., z^{(n)})$. Figure \ref{fig:self_inspiring} provides an example to illustrate the key difference between ToT and SI. In ToT (Figure \ref{fig:self_inspiring}(a)), The new state $N(2)$ at the second path is generated by only considering state $N-1$. The state $N(1)$ is discarded. However, in SI (Figure \ref{fig:self_inspiring}(b)), the new state $N(2)$ is generated based on both $N-1$ and $N(1)$.

The mechanism of SI makes it possible for the agent to analyze different perspectives of the observation of a previous step. For example, an agent for recommending a movie may summarize both the favorite movie director and the favorite movie genre of a user after retrieving the user’s watching history. Next, it can make recommendations from a candidate list by considering both factors. In contrast, previous reasoning methods, such as CoT and ToT, generate the final output based on one single path. Even though ToT samples multiple options at intermediate steps, it only adopts the most confident option and proceeds to the next step. That might be enough for a simple reasoning task. However, recommendation tasks based on textual content require inclusive consideration of different perspectives of available content from both personalized memory and world knowledge.


\begin{algorithm}[t]
\caption{Self-Inspiring Planning}
\begin{algorithmic}[1]
\small
\Require Problem $x$, the current planning path $S=\{z^{(1)}, ..., z^{(m-1)}, s^{(m)}_{j_1}, s^{(m)}_{j_1+1}, ..., s^{(m)}_{t}\}$ at step \( t \), LLM \( p_\theta \), and step limit $T$. Let $\texttt{inspire}(\cdot)$ be the API checking if the planning should explore an alternative reasoning branch.
\While{$t \le T$}
    \State Sample $s^{(m)}_{t+1} = (h^{(m)}_{t+1}, a^{(m)}_{t+1}, o^{(m)}_{t+1}) \sim p_\theta(\cdot | x, S)$
    \If{$h^{(m)}_{t+1}$ is \texttt{"End of Planning"}}
        \State \textbf{break}
    \EndIf
    \State $S' \leftarrow S \cup \{s^{(m)}_{t+1}\}$
    \If{\texttt{inspire}($\{x, S'\}$)}
        \State Sample $s^{(m+1)}_{t+2} \sim p_\theta(\cdot | x, S)$
        \State $S \leftarrow S' \cup \{s^{(m+1)}_{t+2}\}, m \leftarrow m+1, t \leftarrow t+2$
    \Else
        \State $S \leftarrow S', t \leftarrow t+1$
    \EndIf
\EndWhile
\State \Return final response $y \sim p_\theta(\cdot | x, S)$
\end{algorithmic}
\end{algorithm}

\noindent\textbf{Memory} Information stored in memory, including \textit{Personalized Memory} and \textit{World Knowledge}, enables the model to access knowledge beyond what is inherently present in the LLM's parameters. Using the Amazon Reviews dataset as an illustrative example, Personalized Memory includes individualized user information, such as their reviews or ratings for a particular item. World Knowledge consists of two components: the first component is item metadata information, which also falls under the domain-specific knowledge category; the second component involves real-time information that can be accessed through Web search tool. In Figure \ref{fig:example}, information of the product ``Sewak Al-Falah'' retrieved from world knowledge using a Web search tool, aids the reasoning path and ultimately influences the final prediction.

\noindent\textbf{Tool Use}
By empowering LLMs to utilize tools, we can access vastly larger and dynamic knowledge bases, allowing us to tackle complex computational tasks. In RecMind system, we've incorporated three such tools:
\begin{itemize}[leftmargin=*]
\vspace{-2mm}
    \item \textbf{Database Tool}: This tool translates natural language questions into SQL queries. Using this tool, the system can access domain-specific knowledge from memory that's essential for the final prediction. For instance, in the Amazon Reviews dataset, it encompasses personal information such as a user's reviews or ratings for an item, as well as item metadata like the item's description, brand, and price. When the database tool is called, the agent will prompt a question, such as “What is the average rating of product Sewak Al-Falah?", based on the database schema. Next, an LLM is called to transfer the question into an executable SQL query. The output of the SQL execution will then be passed to the agent.
    \vspace{-2mm}

    \item \textbf{Search Tool}: This tool employs a search engine (e.g., Google) to access real-time information. For instance, in the Amazon Reviews dataset, this tool could assist us in obtaining the most recent information about each item. When the Search tool is called, the agent will prompt a question asking for external meta information, which is usually not available in the database, such as “What is the product category of Sewak Al-Falah?". Next, a search engine API will be called to search for the information and return it to the agent.
    \vspace{-2mm}

    \item \textbf{Text Summarization Tool}: This tool helps summarize lengthy texts by invoking a text summarization model from the Hugging Face Hub. For example, within the Amazon Reviews dataset, this tool can produce a summarized description of an item by considering multiple reviews of that specific item from various users. It can generate summarization such as “Most customers think this product is durable and has a good price.", which can be easily used in different recommendation tasks related to the product.

\end{itemize}
Details on the prompts for using the tools and executing self-inspiring are deferred to Section~\ref{supsec:implementation} of the supplementary.
\vspace{-1mm}
\section{Experiments}
\vspace{-1mm}
We evaluate the performance of the RecMind agent in various recommendation scenarios, i.e., rating prediction, sequential recommendation, direct recommendation, explanation generation, review summarization. First, we provide an overview of the datasets and evaluation metrics in Section~\ref{sec:settings} and Section~\ref{sec:compared_methods}. Subsequently, we present the experimental settings and results of RecMind on each recommendation task in Section~\ref{sec::results_precision-oriented} and Section~\ref{sec:results_Explainability}. Next, we study the domain transfer capability of RecMind in Section~\ref{sec:domain_transfer} and how RecMind performs with different foundation LLMs~\ref{sec:ablation_LLM}. In the end, we further explore how the performance of SI in general reasoning tasks in~\ref{sec:general_reasoning} and the running time of RecMind based on SI compared to ToT. The comparison on running time deferred to Section~\ref{sec:running_time} of the supplementary shows that RecMind based on SI takes less inference time than the existing state-of-the-art diverse reasoning method ToT.

\vspace{-2mm}

\subsection{Experimental Settings}
\label{sec:settings}
Following P5 \citep{geng2022recommendation}, we conduct experiments
on the Amazon Reviews~\citep{ni2019justifying} dataset. Since Amazon Reviews contains textual reviews and titles, it provides us the chance to explore how textual contents are utilized in performing recommendation tasks with LLM models.
We evaluate RecMind and baselines on data in Sports $\&$ Outdoors, Beauty, as well as Toys $\&$ Games domains from Amazon Reviews. For a more comprehensive evaluation, we also evaluate the RecMind on Yelp~\citep{geng2022recommendation} dataset. We show the results on the Beauty domain of Amazon Reviews and Telp in Section~\ref{sec::results_precision-oriented} and Section~\ref{sec:results_Explainability}. The results on the Sports and Toys domains of Amazon Reviews are deferred to Section~\ref{sec:results_other_domains} of the supplementary.

To quantitatively evaluate the proposed RecMind across various recommendation tasks, we employ different metrics. For rating prediction, we report Root Mean Square Error (RMSE) and Mean Absolute Error (MAE). In the case of sequential and direct recommendations, we use metrics such as top-\textit{k} Hit Ratio (HR@\textit{k}) and top-\textit{k} Normalized Discounted Cumulative Gain (NDCG@\textit{k}), specifically reporting results on HR@{5,10} and NDCG@{5,10}. In addition, for the assessment of explanation generation, review summarization and conversational recommendation, we use \textit{n}-gram Bilingual Evaluation Understudy (BLEU-\textit{n}) and \textit{n}-gram Recall-Oriented Understudy for Gisting Evaluation (ROUGE-\textit{n}).

\begin{table}[!b]
\centering
\footnotesize
\caption{Performance comparison in rating prediction on Amazon Reviews (Beauty) and Yelp.}
\vspace{-3mm}
\resizebox{\columnwidth}{!}{
\begin{tabular}{lcccc}
\toprule
\multirow{2.5}{*}{Methods} & \multicolumn{2}{c}{\textbf{Beauty}} &  \multicolumn{2}{c}{\textbf{Yelp}} \\
\cmidrule(lr){2-3}\cmidrule(lr){4-5}
 & RMSE  & MAE & RMSE  & MAE  \\
\cmidrule{1-5}
MF & 1.1973 & 0.9461 & \underline{1.2645} & 1.0426 \\
MLP  & 1.3078 & 0.9597 & 1.2951 & 1.0340 \\
AFM & 1.1097 & 0.8815 & \underline{1.2530} & 1.0019 \\
P5 (pre-trained expert,few-shot)  & 1.2982 & 0.8474 & 1.4685  & 1.0054 \\
ChatGPT (zero-shot)  & 1.4173 & 1.1897 & 1.6725 & 1.2359 \\
ChatGPT (few-shot)  & 1.1589 & 0.7327 & 1.4725 & 1.0016 \\
RecMind-CoT (zero-shot)   & 1.2250 & 0.8612 & 1.5302 & 1.1673 \\
RecMind-CoT (few-shot) & 1.1326 & 0.7167 & 1.3925 & 0.9794 \\
RecMind-ToT (BFS, zero-shot)  & 1.1972 & 0.8135 & 1.4956 & 1.0755\\
RecMind-ToT (BFS, few-shot)  & \underline{1.1197} & \underline{0.7059} & 1.3875 & \underline{0.9766} \\
RecMind-ToT (DFS, zero-shot) & 1.2006 & 0.8279 & 1.4937 & 1.1076\\
RecMind-ToT (DFS, few-shot)  & 1.1205 & 0.7103 & 1.3826 & 0.9774 \\
RecMind-SI (zero-shot)  & 1.1894 & 0.7883 & 1.4530 & 1.0009\\
RecMind-SI (few-shot)  & \textbf{1.0756} & \textbf{0.6892} & 1.3674 & \textbf{0.9698} \\
\bottomrule
\end{tabular}
}
\vspace{-1mm}
\label{tab:rating_beauty_yelp}
\end{table}

\begin{table*}[!htbp]
\footnotesize
\centering
\caption{Performance comparison in direct recommendation on Amazon Reviews (Beauty) and Yelp.}
\vspace{-3mm}
\label{tab:direct_beauty_yelp}
\resizebox{1\textwidth}{!}{
\begin{tabular}{lcccccccc} 
\toprule
\multirow{2}{*}{Methods} & \multicolumn{4}{c}{\textbf{Beauty}} & \multicolumn{4}{c}{\textbf{Yelp}}   \\ 
\cmidrule(lr){2-5}\cmidrule(lr){6-9}
                         & HR@5   & NDCG@5 & HR@10  & NDCG@10  & HR@5   & NDCG@5 & HR@10  & NDCG@10   \\ 
\midrule
BPR-MLP                  & {0.1392} & {0.0848} & \textbf{0.2542}& {0.1215} & {0.1876}& {0.1184} &
{0.3066} & 0.1566 \\
ENMF & \textbf{0.1537} & \textbf{0.1124} & \underline{0.2479}& \textbf{0.1453}  & \textbf{0.2235} & \textbf{0.1448} & \textbf{0.3379} & \textbf{0.1851}    \\
P5   (pre-trained expert,few-shot)          & \underline{0.1478} & \underline{0.1003} & {0.2159}& \underline{0.1289}  & \underline{0.2105} & \underline{0.1360} & \underline{0.3182} & \underline{0.1746}    \\
ChatGPT (zero-shot)      & 0.0146 & 0.0107 & 0.0705 & 0.0235 & 0.0479  & 0.0265 & 0.0751 &  0.0326   \\
ChatGPT (few-shot)       & 0.0228 & 0.0157 & 0.0903 & 0.0362  & 0.0512 & 0.0300 & 0.0879 & 0.0412  \\
RecMind-CoT (zero-shot)        & 0.0497 & 0.0325 & 0.1129 & 0.0637 & 0.0992 & 0.0719 & 0.1673 &  0.1170    \\
RecMind-CoT (few-shot)         & 0.0682 & 0.0387 & 0.1345 & 0.0814  & 0.1262 & 0.0897 & 0.1840 & 0.1359    \\
RecMind-ToT (BFS,zero-shot)     & 0.0574 & 0.0439 & 0.1024 & 0.0771  & 0.1032 & 0.0721 & 0.1596 & 0.1273     \\
RecMind-ToT (BFS, few-shot)       & 0.0734 & 0.0402 & 0.1355 & 0.0808  & 0.1649 & 0.0920 & 0.2217 &  0.1503 \\
RecMind-ToT (DFS,zero-shot)    & 0.0564 & 0.0432 & 0.1011 & 0.0751  & 0.1022 & 0.0733 & 0.1587 & 0.1266     \\
RecMind-ToT (DFS, few-shot)       & 0.0705 & 0.0407 & 0.1302 & 0.0812   & 0.1601 & 0.0904 & 0.2079 & 0.1453  \\
RecMind-SI (zero-shot)     & 0.0675 & 0.0524 & 0.1259 & 0.0923  & 0.1055 & 0.0791 & 0.1674 & 0.1293     \\
RecMind-SI (few-shot)      & 0.0915 & 0.0624 & 0.1559 & 0.1063 & 0.1749 & 0.0935  & 0.2451 &  \underline{0.1607} \\ 
\bottomrule
\end{tabular}
}
\vspace{-2mm}
\end{table*}

\begin{table*}[!ht]
\footnotesize
\centering
\caption{Performance comparison in sequential recommendation on Amazon Reviews (Beauty) and Yelp.}
\vspace{-3mm}
\label{tab:sequential_beauty_yelp}
\resizebox{1\textwidth}{!}{
\begin{tabular}{lcccccccc} 
\toprule
\multirow{2}{*}{Methods} & \multicolumn{4}{c}{\textbf{Beauty}} & \multicolumn{4}{c}{\textbf{Yelp}}   \\ 
\cmidrule(lr){2-5}\cmidrule(lr){6-9}
                         & HR@5   & NDCG@5 & HR@10  & NDCG@10  & HR@5   & NDCG@5 & HR@10  & NDCG@10   \\ 
\midrule
S$^3$-Rec                & 0.0387 & 0.0244 & \textbf{0.0647} & 0.0327  & 0.0201 & 0.0123 & 0.0341 & 0.0168 \\
SASRec                   & 0.0401		 & 0.0264	& \underline{0.0643} & 0.0319  & 0.0241 & 0.0175 & 0.0386 & 0.0215 \\
P5 (pre-trained expert,few-shot)& \textbf{0.0459} & \textbf{0.0347} & {0.0603}& \textbf{0.0411} & \textbf{0.0565} & \textbf{0.0389} & 
\textbf{0.0702}  & \textbf{0.0441}  \\
ChatGPT (zero-shot)      & 0.0089 & 0.0053 & 0.0103 & 0.0060 & 0.0102 & 0.0062 & 0.0143 & 0.0089  \\
ChatGPT (few-shot)       & 0.0179 & 0.0124 & 0.0256 & 0.0125 & 0.0217 & 0.0116 & 0.0320 & 0.0165  \\
RecMind-CoT (zero-shot)        & 0.0182 & 0.0139 & 0.0297 & 0.0160 & 0.0368 & 0.0239 & 0.0554 & 0.0316  \\
RecMind-CoT (few-shot)         & 0.0349 & 0.0187 & 0.0486 & 0.0302 & 0.0427 & 0.0305 & 0.0590 & 0.0380  \\
RecMind-ToT (BFS, zero-shot)     & 0.0297 & 0.0172 & 0.0368 & 0.0249 & 0.0379 & 0.0251 & 0.0538 & 0.0322  \\
RecMind-ToT (BFS, few-shot)       & 0.0387 & 0.0235 & 0.0522 & 0.0327 & 0.0447 & 0.0319 & 0.0624 & 0.0337   \\
RecMind-ToT (DFS, zero-shot)     & 0.0299 & 0.0168 & 0.0359 & 0.0241 & 0.0358 & 0.0240 & 0.0519 & 0.0324  \\
RecMind-ToT (DFS, few-shot)       & 0.0365 & 0.0211 & 0.0497 & 0.0355 & 0.0455 & 0.0328 & 0.0622 & 0.0349  \\
RecMind-SI (zero-shot)     & 0.0339 & 0.0200 & 0.0469 & 0.0310 & 0.0396 & 0.0281 & 0.0569 & 0.0340  \\
RecMind-SI (few-shot)      & \underline{0.0415} & \underline{0.0289} & 0.0574 & \underline{0.0375} & \underline{0.0471} & \underline{0.0342} & \underline{0.0635} & \underline{0.0407} \\

\bottomrule
\end{tabular}
}
\vspace{-4mm}
\end{table*}
We use gpt-3.5-turbo-16k \citep{schulman2022chatgpt} as the core large language model in RecMind. To enable the access of RecMind to in-domain knowledge, we store all the review data in a MySQL database, consisting of a table with the product meta information and a table with the interaction history of all the users.  
\vspace{-1mm}
\subsection{Compared Methods} 
\label{sec:compared_methods}
\vspace{-1mm}
We compare the performance of RecMind with the following baselines, including both LLM fine-tuning methods, such as P5~\citep{geng2022recommendation}, and ChatGPT prompting methods~\citep{Liu2023IsCA}. In addition, we implement RecMind with three different planning methods, namely Chain-Of-Thoughts (CoT), Tree-of-Thoughts (ToT)~\citep{Yao2023TreeOT}, and the proposed Self-Inspiring(SI). In summary, the compared methods include:
\begin{itemize}[leftmargin=*]
\vspace{-2mm}
    \item \textbf{P5}~\citep{geng2022recommendation} unifies different recommendation tasks into a shared generative large language model. A collection of personalized prompts has been created for various recommendation-related tasks. All raw data including user-item interactions, user descriptions, item metadata, and users' reviews are transformed into natural language sequences. Subsequently, the large language model is fine-tuned based on these sequences.  In our evaluation, to avoid the influence of factors such as randomness in selecting recommendation candidates, we run the pre-trained P5 model loaded from the Hugging Face repo~\url{https://huggingface.co/makitanikaze/P5} on the same test data we use to evaluate our method.
    \vspace{-2mm}
    \item \textbf{ChatGPT}~\citep{Liu2023IsCA} is a powerful large language model developed by OpenAI. \cite{Liu2023IsCA} constructs a benchmark to evaluate ChatGPT’s performance in different recommendation tasks by designing specific prompts in both zero-shot and few-shot settings. In the zero-shot setting, the LLM is directly prompted for the final prediction, while in the few-shot setting, several in-context examples are provided. 
    We name the ChatGPT baseline in these two settings as \emph{ChatGPT (zero-shot)} and \emph{ChatGPT (few-shot)}.
    \vspace{-2mm}
    \item \textbf{RecMind-CoT}, where the planning is based on ReAct-CoT~\citep{yao2022react}. ReAct is a novel prompt-based paradigm for general task solving. It extends Chain-Of-Thoughts (CoT) \citep{Wei2022ChainOT} to synergize reasoning and acting with external tools. In our experiments, we adopt the same tools we used for the ReAct baseline. We also explore both zero-shot and few-shot for this method and name them as RecMind-CoT (zero-shot) and RecMind-CoT (few-shot).
    \vspace{-2mm}
    \item \textbf{RecMind-ToT}, where the planning is based on Tree-of-Thoughts (ToT)~\citep{Yao2023TreeOT}. ToT enables the exploration of coherent units of thought that serve as intermediate steps toward problem-solving. We implement RecMind-ToT with two strategies in searching among the choices in intermediate steps, which are breadth-first search, named as \emph{RecMind-CoT (BFS, few-shot)} and depth-first search, named as \emph{RecMind-CoT (DFS, few-shot)}.
\end{itemize}
\vspace{-2mm}
In addition to the above methods, we have considered different additional baselines 
for each task. The additional baselines are introduced in corresponding subsections. Details on the prompts for baseline methods are deferred to~\ref{supsec:implementation} of the supplementary.

\subsection{Results on Precision-oriented Recommendation Tasks}
\label{sec::results_precision-oriented}
We first evaluate RecMind and baselines on three precision-oriented recommendation tasks, i.e., rating prediction, sequential recommendation, and direct recommendation. 

\textbf{Rating Prediction.}
Rating prediction is an essential task in recommendation systems that aims to predict the rating that a user would give to a particular item. 
In rating prediction, we further traditional recommendation baselines matrix factorization (MF)~\citep{koren2009matrix}, multi-layer perception (MLP)~\citep{cheng2016wide}, and attentional factorization machines (AFM)~\citep{xiao2017attentional} trained with mean square root loss baselines.
The results of rating prediction on Amazon Reviews (beauty domain) and Yelp are shown in Table~\ref{tab:rating_beauty_yelp}. 
The results show that RecMind with different types of planning mechanisms usually outperforms the fully-trained models for rating prediction tasks. Such improvement mainly stems from the advantage that RecMind has access to both the rating history of the user given to different items and the rating history of the item received from different users in the database. On the other side, fully trained models such as MLP and P5 usually have much higher RMSE, which can be attributed to the over-fitting on the training data. 

\textbf{Direct Recommendation.}
In the scenario of the direct recommendation, RecMind predicts the recommended items from a candidate set of 100 items, where only one candidate is positive. For each test case, we randomly sample 99 candidates from items that the user has never interacted with as negative candidates. 
 Figure \ref{fig:architecture} shows an example of direct recommendation in the beauty domain of Amazon Reviews. For a specific user $\{$userID$\}$ with a list of products, the agent will be prompted, “From the item candidates listed, choose the top 10 items to recommend to the user $\{$userID$\}$ and rank them in order of priority from highest to lowest. Candidates: [`Item List']". We include traditional recommendation method baselines BPR-MLP \citep{cheng2016wide} and ENMF~\citep{chen2020efficient} in this task. The results on direct recommendation are shown in Table~\ref{tab:direct_beauty_yelp}. The results show that fully-trained models such as P5 usually perform better than RecMind. The reason for the performance gap is the long context of the names of 100 candidate items. Specifically, the LLM agent tends to first retrieve information related to items positioned in front of the candidate list. Such positional bias has also been observed in previous works~\citep{Liu2023IsCA}. Table \ref{tab:direct_beauty_yelp} shows that diverse reasoning planning, such as ToT and our proposed SI, benefit in alleviating this issue by gradually filtering out less-possible items. However, it is still hard for LLMs to fully explore the chances of a large candidate set, especially with limitations on prompt context length.

\begin{table*}[!t]
\footnotesize
\centering
\caption{Performance comparison on explanation generation on Amazon Reviews (Beauty) and Yelp.}
\vspace{-3mm}
\label{tab:explanation_beauty_yelp}
\resizebox{1\textwidth}{!}{
\begin{tabular}{lcccccccc} 
\toprule
\multirow{2}{*}{Methods} & \multicolumn{4}{c}{\textbf{Beauty}} & \multicolumn{4}{c}{\textbf{Yelp}}   \\ 
\cmidrule(lr){2-5}\cmidrule(lr){6-9}
                          & BLEU2 & ROGUE1 & ROGUE2 & ROGUEL  & BLEU2 & ROGUE1 & ROGUE2 & ROGUEL \\ 
\midrule
P5   (pre-trained expert,few-shot)          & 0.9783 & \textbf{17.0412} & 1.8962 & \textbf{12.1709} & 1.2784 & \textbf{18.1924} & 2.9517 & \textbf{13.2315}  \\
ChatGPT (zero-shot)      & 0.0359 & 9.7892  & 0.7994 & 5.1215 & 0.0419 & 8.9776 & 0.8549 & 6.1715   \\
ChatGPT (few-shot)       & 1.1766 & 11.8905 & 2.5894 & 5.8920 & 1.1766 & 12.0901 & 3.2170 & 6.7823   \\
RecMind-CoT (zero-shot)        & 0.8985 & 11.0597 & 1.9675 & 7.7471 & 1.1052 & 12.5719 & 2.1941 & 7.7471    \\
RecMind-CoT (few-shot)         & 1.3096 & 12.7987 & 2.7015 & 8.0164 & 1.2759 & 13.9690 & 3.0173 & 9.1081    \\
RecMind-ToT (BFS, zero-shot)         & 1.0279 & 11.1584 & 2.1024 & 7.7026  & 1.1135 & 11.7230 & 2.2355 & 7.7910   \\
RecMind-ToT (BFS, few-shot)       & 1.3054 & 12.8249 & 2.7050 & 8.0596 & \underline{1.2960} & 14.1728 & 3.4539 & 9.6125 \\
RecMind-ToT (DFS, zero-shot)         & 1.0319 & 11.3564 & 2.1416 & 7.7166  & 1.1795 & 11.8433 & 2.2416 & 7.8252   \\
RecMind-ToT (DFS, few-shot)       & \underline{1.3159} & 12.8975 & \underline{2.7125} & 8.1150  & 1.2896 & 14.2201 & \underline{3.6710} & 9.6719 \\
RecMind-SI (zero-shot)         & 1.1589 & 11.6794 & 2.2460 & 7.8974  & 1.1589 & 11.6794 & 2.2460 & 7.8974   \\
RecMind-SI (few-shot)          & \textbf{1.3459} & \underline{13.2560} & \textbf{2.7479} & \underline{8.9614} & \textbf{1.3094} & \underline{14.4220} & \textbf{3.8974} & \underline{9.7125}  \\ 
\bottomrule
\end{tabular}

}
\end{table*}

\textbf{Sequential Recommendation.} For sequential recommendation, the agent takes the names of the user's historically interacted items in order as input. Next, the agent is prompted to predict the title of the next item that the user might interact with. Figure \ref{fig:architecture} shows an example of sequential recommendation in the beauty domain of Amazon Reviews. For a specific user $\{$userID$\}$ with the interaction history in chronological order, the agent will be prompted, “user $\{$userID$\}$ has interacted with the following items in chronological order: [`Item List']. Please recommend the next item that the user might interact with. Choose the top 10 products to recommend in order of priority, from highest to lowest.".  We include traditional recommendation baselines S$^3$-Rec~\citep{zhou2020s3} and SASRec~\citep{kang2018self}. The results in Table~\ref{tab:sequential_beauty_yelp} show that RecMind with Self-Inspiring achieves comparable performance as fully-trained models P5 and S$^3$-Rec. Without diverse planning methods such as tree-of-thoughts and our proposed self-inspiring, LLMs prefer items whose names are semantically similar to those of proceeding items. In contrast, with the help of explicit reasoning methods and access to domain knowledge, RecMind gradually explores helpful information, such as connections of items in the database with other users' interaction history.  
\vspace{-2mm}

\subsection{Results on Explainability-oriented Recommendation Tasks}
\label{sec:results_Explainability}
With the development of NLP techniques on recommendation tasks, recent works~\citep{geng2022recommendation} have started to explore how NLP models can improve the explainability of recommendation systems, such as generating text explanations for a given interaction between a user and an item. In this section, we evaluate the performance of RecMind in two explainability-oriented recommendation tasks, which are explanation generation and review summarization. The results on explanation generation are shown in Table~\ref{tab:explanation_beauty_yelp}. The results on review summarization are deferred to Section~\ref{supsec:review_summarization} of the supplementary.

\textbf{Explanation Generation.} In explanation generation, we assess the performance of RecMind in crafting textual explanations that justify a user's interaction with a specific item. Figure \ref{fig:architecture} shows an example of explanation generation in the beauty domain of Amazon Reviews. The text review given by the user on the given item is taken as the ground truth. The results on explanation generation in Table~\ref{tab:explanation_beauty_yelp} indicates that RecMind, when leveraging self-inspiring techniques, can achieve performance comparable to the fully trained P5 model. This is aided by the in-domain knowledge retrieved from personalized memory, such as reviews from other users on the same item. To better evaluate the quality and rationality of the explanation generated by RecMind and compare the results with baseline models, we perform human evaluation on the generated evaluation. The evaluation details and results are deferred to Section~\ref{supsec:human_evaluation} of the supplementary.
\vspace{-1mm}

\vspace{-1mm}
\subsection{Transfer to Items in Unseen Domains}
\label{sec:domain_transfer}
\vspace{-1mm}
\begin{table}[!htbp]
\centering
\vspace{-1mm}
\caption{Performance on domain transfer. Comparisons are performed on MAE for rating prediction, HR@5 for direct recommendation, and BLEU2 for explanation generation.}
\vspace{-3mm}
\resizebox{\columnwidth}{!}{
\begin{tabular}{ccccc} 
\toprule
Methods                              & Domain          & MAE $\downarrow$ & HR@5 $\uparrow$ & BLEU2 $\uparrow$  \\ 
\midrule
\multirow{2}{*}{P5}                  & Beauty → Toys   & 0.7932                  & 0.0852                       & 1.4326                                                                   \\ 

                                     & Beauty → Sports & 0.7013                  & 0.1007                       & 0.8924                                                                   \\ 
\multirow{2}{*}{ChatGPT}  & Beauty → Toys   & 0.7354                  & 0.0649                       & 1.4416                                                                   \\ 

                                     & Beauty → Sports & 0.6895                  & 0.7210                        & 0.8795                                                                   \\ 

\multirow{2}{*}{RecMind-ToT}      & Beauty → Toys   & 0.6845                  & 0.0841                       & 1.3994                                                                   \\ 

                                     & Beauty → Sports & 0.6457                  & 0.0924                       & 1.0002                                                                   \\ 
\multirow{2}{*}{RecMind-SI } & Beauty → Toys   & 0.6779                  & 0.0902                       & 1.5940                                                                   \\ 

                                     & Beauty → Sports & 0.6245                  & 0.1124                       & 1.0537                                                                   \\
\bottomrule
\end{tabular}
}
\label{tab:domain_transfer}

\end{table}
The advantage of using a large language model as a unified recommendation model is that it can judge the likelihood of any event by expressing the event in natural language. In our experiments in Section~\ref{sec::results_precision-oriented}, we found that RecMind with in-domain few-shot examples achieves much better performance. In this section, we aim to test if the in-domain few-shot examples can generalize to unseen domains, so no parameters need to be trained in such domain transfer experiments. Specifically, we include few-shot examples in the Beauty domain and test the performance of RecMind on rating prediction, direct recommendation, and explanation generation with test data in the Toys and Sports domain. We include ChatGPT prompting baseline and P5 for comparisons. In the few-shot ChatGPT baseline, the user-specific examples included in the prompts are from the Beauty domain. In the P5, the model trained on the Beauty domain is used for evaluation. We evaluate the domain transfer capabilities of all approaches on rating prediction, direct recommendation, and explanation generation. We report the MAE for rating prediction, HR@5 for direct recommendation, and the BLEU2 for explanation in Table~\ref{tab:domain_transfer}. It can be observed that RecMind shows better domain transfer performance compared with the baselines P5 and ChatGPT. In contrast, fine-tuned language model P5 tends to overfit to the domain of the training data.

\vspace{-2mm}
\subsection{Ablation Study on Foundation LLMs}
\label{sec:ablation_LLM}
\begin{figure}[!t]
    \centering
    \includegraphics[width=0.8\columnwidth]{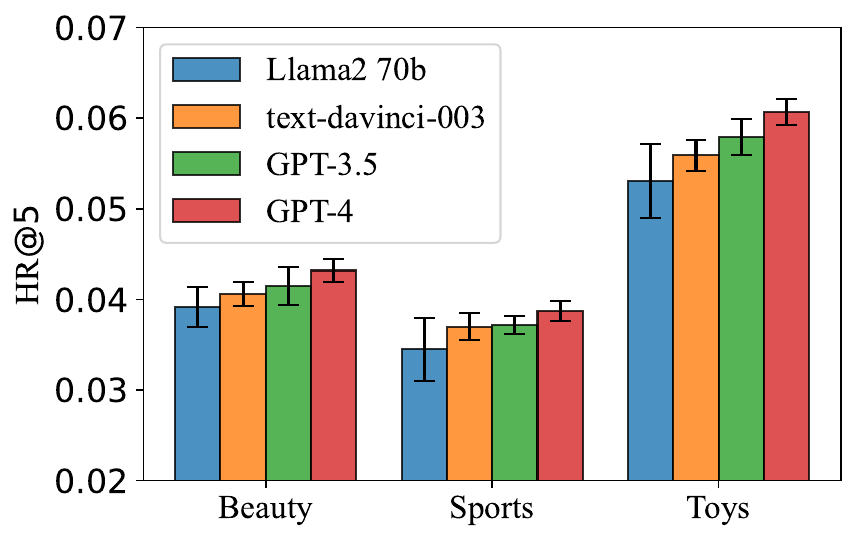}
    \vspace{-3mm}
    \caption{Performance comparison of RecMind-SI with different types of foundation LLMs.} 
    \label{fig:llms_ablation}
    \vspace{-5mm}
\end{figure}

In this section, we study how RecMind performs with different types of foundation LLMs as the controller. We test RecMind-SI using different types of LLMs, including Llama2 70b~\citep{touvron2023llama}, GPT-3.5, text-davinci-003, and GPT-4, for sequential recommendation on three different domains in Amazon Reviews. In each domain, we randomly sample 500 test data for evaluation. We run the evaluation on each model five times and calculate the mean and standard deviation of different runs. The results are shown in Figure~\ref{fig:llms_ablation}. The results show that the performance of RecMind-SI is not sensitive to the selection of Foundation LLMs. Although RecMind-SI with GPT-4 demonstrates enhanced reasoning in addressing complex problems, RecMind-SI with GPT-3.5 can also deliver commendable performance when leveraging the superior capabilities of the RecMind framework. RecMind-SI with Llama2 70b, also achieves pretty good performance. However, due to its limited input context length, the performance with Llama2 has a larger variance.
\subsection{Experiments in general reasoning scenarios}
\label{sec:general_reasoning}
To show that our proposed self-inspiring (SI) method not only outperforms CoT and ToT on recommendation tasks but also on general reasoning scenarios. We evaluate SI on two additional reasoning tasks from [2], which are Game of 24 and Mini Crosswords. We follow the same experimental settings as in ToT [2]. In both tasks, ToT explores the 5 best candidate thoughts at each intermediate step. For a fair comparison, we also set the maximum number of alternative thoughts at each step as 5. We set the maximum number of intermediate steps for the Mini crosswords task to 100 following ToT. GPT-4 backend is used for CoT, ToT, and our SI. The results are shown in Table~\ref{tab:game_of_24} and Table~\ref{tab:CROSSWORDS}. It can be observed that SI outperforms CoT and ToT on both tasks. 
\begin{table}[!htbp]
\centering
\footnotesize
\vspace{-1mm}
\caption{Experiment Results for Game of 24.}
\vspace{-3mm}
\resizebox{0.6\columnwidth}{!}{
\begin{tabular}{cccc}\toprule
Methods  & CoT  & ToT  &  SI   \\
\midrule
Accuracy           & 4 \%           & 74 \%          & 80 \%         \\
\bottomrule
\end{tabular}
}
\label{tab:game_of_24}
\end{table}

\begin{table}[!htbp]
\centering
\vspace{-1mm}
\caption{Experiment Results for Mini Crosswords.}
\vspace{-3mm}
\resizebox{0.75\columnwidth}{!}{
\begin{tabular}{cccc}\toprule
Methods  & CoT  & ToT  &  SI   \\
\midrule
Letter-level Accuracy & 40.6 \% & 78 \% & 81 \% \\
 Word-level Accuracy  & 15.6 \% & 60 \% & 65 \% \\
 Game-level Accuracy  &  1 \%   & 20 \% & 26 \%\\

\bottomrule
\end{tabular}
}
\label{tab:CROSSWORDS}
\end{table}
\vspace{-5mm}
\section{Conclusions}
\vspace{-2mm}
In this work, we propose a novel LLM-powered autonomous agent, RecMind, for various recommendation tasks. The RecMind consists of three major components, i.e., planning, which breaks down a task into smaller sub-tasks; memory, which provides the agent with the capability to retain and recall information over extended periods; and external tools for obtaining relevant extra information from memory that is missing from model weights.  We further propose a novel planning technique self-inspiring, which can integrate the merits of exploring multiple reasoning paths for better planning. We evaluate RecMind across various recommendation tasks, including both precision-oriented tasks and explanability-oriented tasks. The evaluation results show that RecMind with self-inspiring outperforms existing LLM-based recommendation methods in different recommendation tasks and achieves comparable performance to a recent model P5, which is fully trained for the recommendation task. Future works can explore utilizing more external tools in our recommendation agent.

\section*{Limitations}
One major limitation of our work is that more exploration of diverse reasoning paths greatly increases the prompt size, leading to well-known limitations of LLMs in long contexts and position bias. A future direction could be implementing a summarization step for historical paths, which might not only condense the long context but also potentially remove some of the noise in historical paths. In addition, only a small number of external tools are adopted in our current implementation. 

\section*{Ethical Concerns and Broader Impacts}
All experiments in our papers are performed on two widely used recommendation datasets, which are Amazon Reviews~\cite{ni2019justifying} and Yelp~\cite{ni2019justifying}. To protect users' privacy, both datasets adopt anonymous user IDs to represent user identity. We follow the terms of use for both datasets and only use the datasets for academic purposes. 
The LLM-based recommendation system proposed in this work has the potential to influence consumer behavior and preferences. In addition, we have tested the method on top of different LLM models, including online and offline models, to avoid potential biases in pre-trained LLMs such as ChatGpt~\cite{schulman2022chatgpt}.

\bibliography{anthology,custom}
\bibliographystyle{acl_natbib}
\clearpage
\appendix
\section{Additional Implementation Details}
\label{supsec:implementation}
\noindent\textbf{Tool Descriptions in Agent Prompt}
To enable the LLM-based Agent to utilize external tools, the LLM Agent will be prompted an instruction with descriptions on different tools. The prompt is formulated as:

\noindent Perform a recommendation task with interleaving Thought, Action, and Observation steps. Thought can reason about the current situation, and Action can be the following types: 
\begin{itemize}[leftmargin=*]
    \item  SQL Tool: ``SQL \{question\}, which aims to search for the answer to a question from the database. You can only put forward questions based on the available information in the database. Available information and schema of the database is provided in \{database\_info\}.''
    \item Text Summarization Tool: ``Summarize \{content\}, which condenses extensive text into a shorter version while retaining the core information and meaning by using a pre-trained text summarization model.''
    \item Search Tool: ``Search \{question\}, which formulates a search query for Google search engine based on the question. This tool can be used to search for information that is unavailable in the database."
    \item Finish: ``Finish \{answer\}, which returns the answer and finishes the task.''
\end{itemize}

\noindent\textbf{Search Tool Prompt} In the search tool, we use \url{SerpApi.com} as our Google search API. Since the output of the search API is in a structured JSON format, we use the same LLM model of the agent to convert the output to a text response and then return it to the LLM agent. The prompt we use is ``Your mission is to convert the Google search result \{search\_result\} from search engine to meaningful sentences, which can be a response to question \{question\}.''         

\noindent\textbf{SQL Tool Prompt} In the SQL tool, we use the same LLM model of the agent to convert the question to an SQL query. The prompt we use in this text-to-SQL process is ``Your mission is to convert SQL query from given \{question\}. The information about the tables in the database is \{database\_info\}. Only output the SQL query.'' Next, the obtained SQL query will be executed. Similar to the search tool, the output will then be converted to a text response to the question and returned to the LLM agent. The prompt we use to convert the output is ``Your mission is to convert SQL query execution results to meaningful sentences, which should be the answer to the question \{question\}. The query generated for this question is \{sql\_query\}. Here is the database result: \{sql\_result\}''

\noindent\textbf{Self-Inspiring Prompt} In the implementation of self-inspiring, the same LLM model of the agent is used to decide whether another thought is necessary given the task and previously explored steps. The prompt for this request is 
``You are given multi-step problem-solving steps towards finishing the task \{task\}. The previous steps are \{previous\_steps\}. You already have the thought, action, and observation in the current step \{current\_step\}.
Your mission is to decide if there is an alternative thought in the current step that can help finish this task following the previous steps. If there is, directly output the thought. If not, please respond \{empty\_response\}.''

For  ChatGPT (zero-shot) and ChatGPT (few-shot), we use the exact same prompt templates from~\citep{Liu2023IsCA}. We will attach the prompt templates for all baseline methods in the appendix of the revised version of our paper. 
We follow~\citep{yao2022react} to design the prompt for CoT. The prompt is ``Solve a recommendation task with interleaving Thought, Action, and Observation steps.''
We follow \citep{Yao2023TreeOT} to design the prompts for ToT. In addition to the general instruction, ``Solve a recommendation task with interleaving Thought, Action, and Observation steps", we also designed prompts for thought-sampling and decision-making.
The thought sampling prompt is ``Given the previous \{previous\_steps\}, list five possible thoughts for the next step towards finishing the task \{task\}.'' 
The decision-making prompt is ``Given an instruction and several choices, decide which choice is most promising. Your instruction is \{task\_sepcific\_instruction\}. Your available options are \{option\_list\}. Analyze each choice, then conclude in the last line, `The best choice is \{s\}', where s is the integer id of the choice.''
\section{Additional Experiment Results}
\subsection{Results on Review Summarization}
\label{supsec:review_summarization}
In the review summarization task, we evaluate the performance of RecMind in summarizing review comments to shorter review titles. We filter out test data with automatically generated review titles such as 'Five Stars'. Figure \ref{fig:architecture} shows an example of review summarization in the beauty domain of Amazon Reviews. The results of the review summarization on Amazon Reviews are shown in Table~\ref{tab:summarization_beauty_yelp}. The result shows that RecMind agent performs better that recent LLM such as ChatGPT. However, RecMind does not outperform P5 regarding the review summarization. This performans comes from the advantage of P5 which fully trained model towards optimizaing the review summarization task. In contrast, GPT-based models, such as RecMind, usually prioritize generating summaries after deeply understanding the reviews. 
\begin{table}[!htbp]
\footnotesize
\centering

\caption{Performance comparison on review summarization on Amazon Reviews (Beauty).}
\vspace{-3mm}
\label{tab:summarization_beauty_yelp}
\resizebox{\columnwidth}{!}{
\begin{tabular}{lcccc} 
\toprule
\multirow{2}{*}{Methods} & \multicolumn{4}{c}{\textbf{Beauty}} \\ 
\cmidrule(lr){2-5}
                          & BLEU2 & ROGUE1 & ROGUE2 & ROGUEL  \\ 
\midrule
P5   (pre-trained expert,few-shot)          & \textbf{2.0357} & \textbf{8.3079} & \textbf{1.5892} & \textbf{7.4820}  \\
ChatGPT (zero-shot)      & 0.6532 & 3.8579 & 0.3059 & 3.3552 \\
ChatGPT (few-shot)       & 0.9137 & 4.0179 & 0.4179 & 3.6790  \\
RecMind-CoT (zero-shot)        & 1.3596 & 5.0279 & 0.7156 & 4.7689 \\
RecMind-CoT (few-shot)         & 1.3786 & 5.5397 & 0.8456 & 4.8024 \\
RecMind-ToT (BFS, zero-shot)     & 1.3592 & 5.1103 & 0.7596 & 4.8069 \\
RecMind-ToT (BFS, few-shot)      & 1.3737 & 5.4187 & 0.8254 & 4.8157  \\
RecMind-ToT (DFS, zero-shot)     & 1.3614 & 5.1435 & 0.7749 & 4.7985 \\
RecMind-ToT (DFS, few-shot)      & 1.3798 & 5.5794 & 0.8351 & 4.8976 \\
RecMind-SI (zero-shot)     & 1.3688 & 5.4579 & 0.8974 & 4.9746 \\
RecMind-SI (few-shot)      & \underline{1.4014} & \underline{6.0354} & \underline{1.0128} & \underline{5.5716} \\
\bottomrule
\end{tabular}
}
\end{table}

\begin{table}[!b]
\centering
\vspace{-4mm}

\caption{Human evaluation results on explanation generation. }
\vspace{-3mm}
\resizebox{0.7\columnwidth}{!}{
\begin{tabular}{ccccc}
\toprule
\multirow{2}{*}{Methods} & \multicolumn{3}{c}{Evaluator} & \multirow{2}{*}{Average} \\ \cmidrule{2-4}
                         & Eva\_1   & Eva\_2   & Eva\_3  &                                   \\ 
\midrule
Ground Truth             & 0.12     & 0.13     & 0.22    & 0.157                             \\
P5                       & 0.02     & 0.06     & 0.03    & 0.037                             \\
ChatGPT                  & 0.15     & 0.23     & 0.18    & 0.187                             \\
RecMind-ToT             & 0.29     & 0.28     & 0.25    & \underline{0.273}                             \\
RecMind-SI              & 0.42     & 0.30     & 0.32    & \textbf{0.347 }                            \\ 
\bottomrule
\end{tabular}
}
\label{tab:human}
\end{table}
\vspace{-5mm}

\subsection{Human Evaluation}
\label{supsec:human_evaluation}
In this section, we leverage human evaluation to assess the quality and rationality of the explanation generated by RecMind. Three human evaluators (Eva\_1, Eva\_2, Eva\_3) are asked to rank the explanations generated by P5, few-shot ChatGPT, few-shot RecMind with tree-of-thoughts, few-shot RecMind with self-inspiring and the ground truth on 100 test data. We show the top-1 ratios on results generated by different methods in Table~\ref{tab:human} for each evaluator. The top-1 ratio indicates the proportion of test data where the given method ranks first compared to other methods based on each annotator's selection. We also calculate the average top-1 ratios of all three evaluators on results generated by each method. Although annotators may have individual subjectivity, evaluations by different evaluators consistently show that the few-shot RecMind based on self-inspiring, i.e., RecMind-SI yields the most satisfactory results. 
\subsection{Running Time Analysis}
\label{sec:running_time}
In this section, we provide a running time comparison between our proposed reasoning method SI and previous reasoning methods for the recommendation agent. We run RecMind with CoT, ToT, and SI on 100 randomly sampled test data from the Beauty domain of Amazon Reviews and calculate the average running time. We use GPT-3.5 as the base model. The results in Table~\ref{tab:running_time} show that our proposed self-inspiring can not only improve the performance of the LLM-powered agent but also take less inference time than the existing state-of-the-art diverse reasoning method ToT. Such merit mainly stems from the fact that SI only explores alternative options at an intermediate step when it recognizes that the explored options at that step are not good enough. In contrast, ToT directly samples multiple options for exploration, which can lead to a waste of time.

\begin{table}[!htbp]
\centering
\vspace{-1mm}
\caption{Average Running Time of RecMind with Different Reasoning Methods.}
\vspace{-3mm}
\resizebox{0.8\columnwidth}{!}{
\begin{tabular}{cccc}\toprule
Methods  & CoT  & ToT  &  SI   \\
\midrule
Average Running Time (s) & 18.9 s & 53.2 s & 29.7 s \\
\bottomrule
\end{tabular}
}
\label{tab:running_time}
\end{table}
\begin{table}[!b]
\centering
\footnotesize
\caption{Performance comparison in rating prediction on Sports and Toys domains of Amazon Reviews.}
\vspace{-3mm}
\resizebox{1\columnwidth}{!}{

\begin{tabular}{lcccc}
\toprule
\multirow{2.5}{*}{Methods} & \multicolumn{2}{c}{\textbf{Sports}} &  \multicolumn{2}{c}{\textbf{Toys}} \\
\cmidrule(lr){2-3}\cmidrule(lr){4-5}
 & RMSE  & MAE & RMSE  & MAE  \\
\cmidrule{1-5}
MF &\underline{1.0274} & 0.7975 &\underline{1.0193} & 0.8024 \\
MLP  &1.1277 & 0.7626 &1.1215 & 0.8097 \\
P5 (pre-trained expert,few-shot)  &1.0534 & 0.6784 &1.0625 & 0.7134 \\
ChatGPT (zero-shot)  &1.2723 & 1.0637 &1.3213 & 1.0117 \\
ChatGPT (few-shot)  &1.0929 & 0.6957 &1.0519 & 0.7047 \\
RecMind-CoT (zero-shot)   &1.1490 & 0.8042 &1.1680 & 0.8232 \\
RecMind-CoT (few-shot) &1.0325 & 0.6446 &1.0403 & 0.6905 \\
RecMind-ToT (BFS, zero-shot)  &1.1322 & 0.8014 &1.1559 & 0.8164\\
RecMind-ToT (BFS, few-shot)  &1.0307 & \underline{0.6289} &1.0279 & 0.6823 \\
RecMind-ToT (DFS, zero-shot)  &1.1366 & 0.8021 &1.1537 & 0.8155\\
RecMind-ToT (DFS, few-shot)  &1.0545 & 0.6433 &1.0196 & \underline{0.6801} \\
RecMind-SI (zero-shot)  &1.1230 & 0.7913 &1.1412 & 0.8103
\\
RecMind-SI (few-shot)  &\textbf{1.0124} & \textbf{0.6122} &\textbf{1.0086} & \textbf{0.6712} \\
\bottomrule
\end{tabular}
}
\label{tab:rating_sports_toys}
\end{table}
\vspace{-5mm}
\subsection{Results on Sports and Toys Domains in Amazon Reviews}
\label{sec:results_other_domains}
In this section, we provide additional experiment results of RecMind and all compared methods on the Sports domain and Toys domain in Amazon Reviews. The results in rating prediction on the Sports and Toys domains of Amazon Reviews are shown in Table~\ref{tab:rating_sports_toys}. The results in the direct recommendation and sequential recommendation on the Sports domain of Amazon Reviews are shown in Table~\ref{tab:direct_sequential_sports}. The results in the direct recommendation and sequential recommendation on the Toys domain of Amazon Reviews are shown in Table~\ref{tab:direct_sequential_toys}. The results in text summarization and explanation generation on the Sports domain of Amazon Reviews are shown in Table~\ref{tab:summarization_explanation_sports}. The results in text summarization and explanation generation on the Toys domain of Amazon Reviews are shown in Table~\ref{tab:summarization_explanation_toys}. As indicated in the experimental results, RecMind also performs well in different recommendation tasks on data from other domains of Amazon Reviews.

\begin{table}[!htbp]
\centering
\caption{Performance comparison in direct recommendation and sequential recommendation
on Sports domain of Amazon Reviews.}
\vspace{-3mm}
\label{tab:direct_sequential_sports}
\resizebox{1\columnwidth}{!}{
\begin{tabular}{lcccc} 
\toprule
\multirow{2}{*}{Methods}  & \multicolumn{4}{c}{\textbf{Sports}}  \\ 
\cmidrule(lr){2-5}
                           & HR@5 & NDCG@5 & HR@10 & NDCG@10         \\ 
\midrule
\multicolumn{5}{c}{Direct Recommendation}\\
\midrule
BPR-MLP                    & \underline{0.1520} & \underline{0.0927} & \textbf{0.2671} &0.1296   \\
P5   (pre-trained expert,few-shot)             & \textbf{0.1765} & \textbf{0.1196} & \underline{0.2235}& \textbf{0.1325}      \\
ChatGPT (zero-shot)     & 0.0376 &0.0317 & 0.0902 & 0.0459      \\
ChatGPT (few-shot)      & 0.0388 &0.0267 & 0.1003 & 0.0502      \\
RecMind-CoT (zero-shot)        & 0.0607 &0.0435 & 0.1259 & 0.0757      \\
RecMind-CoT (few-shot)        & 0.0782 &0.0527 & 0.1475 & 0.1034      \\
RecMind-ToT (BFS, zero-shot)    & 0.0741 &0.0512 & 0.1320 & 0.1054      \\
RecMind-ToT (BFS, few-shot)      & 0.0874 &0.0542 & 0.1475 & 0.1218      \\
RecMind-ToT (DFS, zero-shot)    & 0.0759 &0.0519 & 0.1320 & 0.1079      \\
RecMind-ToT (DFS, few-shot)      & 0.0815 &0.0557 & 0.1412 & 0.1272      \\
RecMind-SI (zero-shot)    & 0.0835 &0.0684 & 0.1379 & 0.1103      \\
RecMind-SI (few-shot)     & 0.1115 &0.0814 & 0.1769 & \underline{0.1303}  \\ 
\midrule
\multicolumn{5}{c}{Sequential Recommendation}\\
\midrule
S$^3$-Rec                  &0.0251 & 0.0161 & 0.0385 & 0.0204     \\
P5   (pre-trained expert,few-shot)           & \underline{0.0357} & \textbf{0.0289} & 0.0416 & \textbf{0.0324}  \\
ChatGPT (zero-shot)        & 0.0039 &0.0008 & 0.0051 & 0.0008      \\
ChatGPT (few-shot)         & 0.0130 &0.0075 & 0.0207 & 0.0070      \\
RecMind-CoT (zero-shot)           & 0.0135 &0.0090 & 0.0248 & 0.0105      \\
RecMind-CoT (few-shot)           & 0.0300 &0.0138 & 0.0437 & 0.0247      \\
RecMind-ToT (BFS, zero-shot)        & 0.0205 &0.0134 & 0.0319 & 0.0243      \\
RecMind-ToT (BFS, few-shot)         & 0.0338 &0.0186 & \underline{0.0473}  & 0.0272  \\
RecMind-ToT (DFS, zero-shot)        & 0.0218 &0.0130 & 0.0336 & 0.0238      \\
RecMind-ToT (DFS, few-shot)         & 0.0316 &0.0162 & 0.0448 & 0.0260      \\
RecMind-SI (zero-shot)        & 0.0290 &0.0151 & 0.0420 & 0.0255      \\
RecMind-SI (few-shot)      & \textbf{0.0366} & \underline{0.0240} & \textbf{0.0525} & \underline{0.0320}   \\
\bottomrule
\end{tabular}
}
\end{table}

\begin{table}[!htbp]
\centering

\caption{Performance comparison in direct recommendation and sequential recommendation
on Toys domain of Amazon Reviews.}
\vspace{-3mm}
\label{tab:direct_sequential_toys}
\resizebox{1\columnwidth}{!}{
\begin{tabular}{lcccc} 
\toprule
\multirow{2}{*}{Methods}  & \multicolumn{4}{c}{\textbf{Toys}}  \\ 
\cmidrule(lr){2-5}
                           & HR@5 & NDCG@5 & HR@10 & NDCG@10         \\ 
\midrule
\multicolumn{5}{c}{Direct Recommendation}\\
\midrule
BPR-MLP                   & \underline{0.1142} & \underline{0.0688} & \textbf{0.2077} & \textbf{0.0988}   \\
P5   (pre-trained,few-shot)             & \textbf{0.1278} & \textbf{0.0743} & \underline{0.1859}& 0.1089      \\
ChatGPT (zero-shot)     & 0.0114 & 0.0075 & 0.0638 & 0.0191      \\
ChatGPT (few-shot)      & 0.0130 & 0.0059 & 0.0805 & 0.0270      \\
RecMind-CoT (zero-shot)        & 0.0399 & 0.0233 & 0.1031 & 0.0542      \\
RecMind-CoT (few-shot)        & 0.0580 & 0.0295 & 0.1247 & 0.0719      \\
RecMind-ToT (BFS,zero-shot)     & 0.0496 & 0.0297 & 0.1079 & 0.0697      \\
RecMind-ToT (BFS, few-shot)      & 0.0636 & 0.0300 & 0.1257 & 0.0813      \\
RecMind-ToT (DFS,zero-shot)     & 0.0510 & 0.0301 & 0.1094 & 0.0712      \\
RecMind-ToT (DFS, few-shot)      & 0.0603 & 0.0315 & 0.1204 & 0.0817      \\
RecMind-SI (zero-shot)     & 0.0577 & 0.0432 & 0.1161 & 0.0828      \\
RecMind-SI (few-shot)      & 0.0813 & 0.0532 & 0.1461 & \underline{0.0998}  \\ 
\midrule
\multicolumn{5}{c}{Sequential Recommendation}\\
\midrule
S$^3$-Rec                  &0.0443 & 0.0294 & \underline{0.0700} &0.0376      \\
P5   (pre-trained,few-shot)           &  \textbf{0.0612} & \textbf{0.0524} & \textbf{0.0702} & \textbf{0.0569} \\
ChatGPT (zero-shot)        & 0.0192 & 0.0158 & 0.0212 & 0.0165   \\
ChatGPT (few-shot)         & 0.0282 & 0.0231 & 0.0367 & 0.0230   \\
RecMind-CoT (zero-shot)           & 0.0285 & 0.0246 & 0.0408 & 0.0265   \\
RecMind-CoT (few-shot)           & 0.0452 & 0.0294 & 0.0597 & 0.0407   \\
RecMind-ToT (BFS,zero-shot)     & 0.0399 & 0.0287 & 0.0495 & 0.0359      \\
RecMind-ToT (BFS, few-shot)         & 0.0490 & 0.0342 & 0.0633 & 0.0432  \\
RecMind-ToT (DFS,zero-shot)     & 0.0412 & 0.0295 & 0.0507 & 0.0376     \\
RecMind-ToT (DFS, few-shot)         & 0.0468 & 0.0318 & 0.0608 & 0.0420   \\
RecMind-SI (zero-shot)        & 0.0442 & 0.0307 & 0.0580 & 0.0415   \\
RecMind-SI (few-shot)      & \underline{0.0518} & \underline{0.0396} & 0.0685 & \underline{0.0480}    \\
\bottomrule
\end{tabular}
}
\end{table}
\newpage
\begin{table}[!htbp]
\centering
\caption{Performance comparison on review summarization and explanation generation on Sports domain of Amazon Reviews.}
\vspace{-3mm}
\label{tab:summarization_explanation_sports}
\resizebox{1\columnwidth}{!}{
\begin{tabular}{lcccc} 
\toprule
\multirow{2}{*}{Methods} & \multicolumn{4}{c}{\textbf{Sports}}        \\ 
\cmidrule(lr){2-5}
                          & BLEU2 & ROGUE1 & ROGUE2 & ROGUEL   \\ 
\midrule
\multicolumn{5}{c}{Review Summarization}\\
\midrule
P5   (pre-trained expert,few-shot)& \textbf{2.5874} & \textbf{11.8971} & \textbf{3.0257} & \textbf{10.5472}    \\
ChatGPT (zero-shot)               & 0.9024 & 5.7402  & 1.2493 & 3.6791   \\
ChatGPT (few-shot)                & 1.2579 & 6.3190  & 1.4257 & 3.8912   \\
RecMind-CoT (zero-shot)           & 1.5840 & 6.5310  & 1.4390 & 5.0140   \\
RecMind-CoT (few-shot)            & 1.6014 & 6.7125  & 1.5479 & 5.2175   \\
RecMind-ToT (BFS, zero-shot)      & 1.5940 & 6.5872  & 1.4780 & 5.1566   \\
RecMind-ToT (BFS, few-shot)       & 1.7125 & 6.7986  & 1.5724 & 5.3794   \\
RecMind-ToT (DFS, zero-shot)      & 1.5874 & 6.5531  & 1.4726 & 5.1530   \\
RecMind-ToT (DFS, few-shot)       & 1.6542 & 6.6540  & 1.5639 & 5.2960   \\
RecMind-SI (zero-shot)            & 1.6120 & 6.6259  & 1.5029 & 5.1891   \\
RecMind-SI (few-shot)             & \underline{1.7388} & \underline{6.8130}  & \underline{1.6217} & \underline{5.5632} \\

\midrule
\multicolumn{5}{c}{Explanation Generation}\\
\midrule
P5   (pre-trained expert,few-shot)         & 1.1412 & \textbf{14.0329} & 2.1279 & \textbf{11.1894} \\
ChatGPT (zero-shot)            & 0.0611 & 7.2892  & 0.9921 & 5.6923    \\
ChatGPT (few-shot)             & 1.2358 & 9.6405  & 2.8723 & 6.2824    \\
RecMind-CoT (zero-shot)        & 0.9687 & 8.3097  & 2.1320 & 7.1427    \\
RecMind-CoT (few-shot)         & 1.3874 & 11.0487 & 3.0216 & 8.1146    \\
RecMind-ToT (BFS, zero-shot)   & 1.1032 & 8.9895  & 2.3810 & 7.8419   \\
RecMind-ToT (BFS, few-shot)    & 1.3765 & 11.5749 & 2.8023 & 8.4256     \\
RecMind-ToT (DFS, zero-shot)   & 1.1345 & 9.0957  & 2.4866 & 7.9965   \\
RecMind-ToT (DFS, few-shot)    & \underline{1.4018} & 11.6475 & \underline{3.0107} & 8.6032  \\
RecMind-SI (zero-shot)         & 1.2374 & 9.4294  & 2.5405 & 8.2120   \\
RecMind-SI (few-shot)          & \textbf{1.4287} & \underline{12.0060} & \textbf{3.0481} & \underline{9.5812}    \\ 
\bottomrule
\end{tabular}

}
\end{table}

\begin{table}[!htbp]
\centering
\caption{Performance comparison in review summarization and explanation generation on Toys domain in Amazon Reviews.}
\vspace{-3mm}
\label{tab:summarization_explanation_toys}
\resizebox{1\columnwidth}{!}{
\begin{tabular}{lcccc} 
\toprule
\multirow{2}{*}{Methods} & \multicolumn{4}{c}{\textbf{Toys}}        \\ 
\cmidrule(lr){2-5}
                          & BLEU2 & ROGUE1 & ROGUE2 & ROGUEL   \\ 
\midrule
\multicolumn{5}{c}{Review Summarization}\\
\midrule
P5   (pre-trained expert,few-shot)          & \textbf{1.8760} & \textbf{9.0351} & \textbf{1.5230} & \textbf{8.1746}    \\
ChatGPT (zero-shot)            & 0.5941 & 4.4571 & 0.4052 & 4.0612   \\
ChatGPT (few-shot)             & 0.8420 & 4.8179 & 0.3178 & 4.2889   \\
RecMind-CoT (zero-shot)        & 1.1579 & 5.7276 & 0.7158 & 5.5691   \\
RecMind-CoT (few-shot)         & 1.2394 & 6.3395 & 0.9453 & 5.8123   \\
RecMind-ToT (BFS, zero-shot)   & 1.1603 & 5.9315 & 0.8259 & 5.4930   \\
RecMind-ToT (BFS, few-shot)    & 1.2668 & 6.3186 & 0.9251 & 5.6159   \\
RecMind-ToT (DFS, zero-shot)   & 1.1725 & 6.0014 & 0.8551 & 5.5012   \\
RecMind-ToT (DFS, few-shot)    & 1.2515 & 6.2791 & 0.9356 & 5.5976   \\
RecMind-SI (zero-shot)         & 1.1897 & 6.2578 & 0.8976 & 5.8724   \\
RecMind-SI (few-shot)          & \underline{1.2974} & \underline{6.8352} & \underline{1.1125} & \underline{6.2718} \\

\midrule
\multicolumn{5}{c}{Explanation Generation}\\
\midrule
P5   (pre-trained expert,few-shot)          & 2.2850 & \textbf{15.0416} & \underline{3.6798} & \textbf{12.1065}   \\
ChatGPT (zero-shot)            & 0.1379 & 9.7892  & 1.5416 & 5.3158   \\
ChatGPT (few-shot)             & 2.0169 & 11.8905 & 3.2049 & 6.2689   \\
RecMind-CoT (zero-shot)        & 2.1354 & 11.0597 & 2.1590 & 7.1445   \\
RecMind-CoT (few-shot)         & 2.4079 & 12.7987 & 3.5146 & 7.4153   \\
RecMind-ToT (BFS, zero-shot)   & 2.1930 & 11.2874 & 2.1782 & 7.1854    \\
RecMind-ToT (BFS, few-shot)    & \underline{2.4565} & 12.8249 & 3.6327 & 7.6234   \\
RecMind-ToT (DFS, zero-shot)   & 2.1658 & 11.2802 & 2.1770 & 7.1809    \\
RecMind-ToT (DFS, few-shot)    & 2.4152 & 12.8975 & 3.6079 & 7.7112    \\
RecMind-SI (zero-shot)         & 2.2740 & 11.6794 & 2.2460 & 7.2536    \\
RecMind-SI (few-shot)          & \textbf{2.4674} & \underline{13.2560} & \textbf{3.6920} & \underline{7.9987}    \\ 
\bottomrule
\end{tabular}

}
\end{table}

\end{document}